*Review Article*

# Managing Innovation in Technical Education: Revisiting the Developmental Strategies of Politeknik Brunei

**Bashir Ahmed Bhuiyan[1], Mohammad Shahansha Molla[1,*] and Masud Alam[2]**

[1]Leading University, Bangladesh
bashir.dba1970@gmail.com; shahansha06@yahoo.com
[2]Shahjalal University of Science and Technology, Bangladesh
masudalam-eco@sust.edu
*Correspondence: shahansha06@yahoo.com

**Abstract:** Present study aims at exploring the strategies for managing innovation in technical education by using blended learning philosophy and practices with special reference to Politeknik Brunei. Based on literature review and desk research, the study found out salient characteristics, explored constraining factors, elicited strategies of Politeknik Brunei, and suggested some options and a framework for innovations management and development of effective blended teaching and learning. The limiting factors identified are: the unwillingness of the top-level management, lack of structural support, dearth of readiness of the stakeholders, the gap between teachers' expectations and changed students' characteristics, and blended teaching myopia on the way of effective application of blended learning strategies. Notable suggestions for strategic development are developing wide-angle vision and self-renewal processes, analyzing the environment for needs determination. Clarity of purpose and tasks, technological adaptability, data-driven decision making, prompt feedback, flipped classroom, and development of learning clusters are other dimensions that may go a long way toward innovating teaching-learning and the overall development of an academic institution. Finally, the study suggested important guidelines for applying the strategies and proposed framework for quality blended learning and managing innovations in technical education.

## 1. Introduction

Managing innovation in teaching and learning plays an important role in the digitalization age that offers students the opportunity to learn new skills or improve existing ones (Rugova, 2021). Blended learning has been considered one of the effective student-centric learning strategies that prepare learners for success in a rapidly changing innovative environment (Garrison & Kanuka, 2004; Medina, 2018). This enables critical thinking, demands technological dexterity, and develops flexible intelligence to drive through and thrive by the changes (Han & Ellis, 2019). Active learning contributes to improving learning, supports knowledge retention, and raises achievements throughout the learning processes. This learning strategy inspires engagements, acknowledges the reality of environments, and emphasizes the multiplicity of learners requirements whereby learners learn through living experiences facilitated by information and communication technology and access through 24/7 smartphones, laptops and tablets etc (Ngigi & Obura, 2019; Serrano, Dea-Ayuela, Gonzalez-Burgos, Serrano-Gil, & Lalatsa, 2019). Student engagement and allowing freedom help them accomplish their dreams and develop themselves through good relationships with multiorbital influencers, the professional qualities of the teachers, and innovative teaching-learning methods. The goals for expected students' engagement can be achieved through the support of teachers, the school, and the harmonious relationship with peers whereby maintaining contemporariness and applying competency and proficiency-based systems, students move through the flexible pace of learning go a long way through innovations and changes in the teaching-learning environment and technological innovations(Mamun, Bhuiyan, Rab, & Islam, 2000; Serrano et al., 2019). However, through the paradigm shift in learning philosophy, execution of learning processes, and control of learning outcomes, the blended design of learning focuses on designing, executing, and controlling learning outcomes with the goal of accelerated learning. This study aims to give a theoretical overview and models on blended learning, provide characteristics of blended learning, practices in the politeknik Brunei, explore the limitations and gaps in the application, and finally provides alternatives and lessons for improvement of learning and applying the strategies for development in the competitive environment

## 2. Literature Review on Innovation, Blended Learning and Sustainable Competitive Advantage of the Academic Institutions

Searching for, responding to, and exploiting opportunities is one of the important functions of innovation. Innovations are the essential practices that create a sustainable competitive advantage for academic institutions (Bhuiyan, 2001; Molla, Ibrahim, & Ishak, 2019). An educational institution should develop strategies by considering its history, analyzing its present status, and orienting towards a future that captures multiple learning objectives that students may obtain through blended learning and teaching systems(Bhuiyan & Ahmed, 2011). There is no unanimous definition regarding blended learning approaches in education(Vaughan, 2019). A variety of definitions cover different aspects of instructions. Combining online delivery with the best live classroom interactions and blended teaching provides personalized learning, thoughtful reflection, and differentiated instructions across diverse learners. This strategy effectively integrates the information and communication technologies to design and enhance teaching and learning experiences where most learners' centric participation is ensured either in face-to-face or distance mode of learning and intentionally capitalizes on the best features of each learning modality. Combining a mix of delivery modes, teaching styles, and learning approaches to blending strategy delivers better students' understandings and outcomes and efficient teaching and course management practices(Bath & Bourke, 2010). Sustainable competitive advantage and sustainable development are linked to each other that assure long-term survival, financial success, and competitive presence of an enterprise or an academic institution (Molla, 2021; Molla, Hasan, Miraz, Azim, & Hossain, 2021; Molla, Ibrahim, & Ishak, 2019).

In the ensuing age of learning and teaching, higher education goes beyond the surface conclusion that students are applying interactive media, instead of from the students' usage, new horizons are emerging —robust engagement and learning, thinking styles, and new literacies— from which institutions should rethink how we view the creation, sharing and mastery of knowledge (Dede, 2009). Blended learning refers to different practices and strategies by which "students learn in part online, with some element of control over the time, place, path, or pace of their learning; in part in a brick-and-mortar location away from home; [and] the modalities along a students` learning path are connected to provide an integrated learning experience.[1]" Blended learning usually use strategic integration of technology to support and adjust instructional methods to suit the students' evolving needs by teachers and to shift contents and instructions for accelerating learning, leveraging talents through monitoring progress, analyzing real-time learning data and controlling time, place and pace by students over the completion of their study and works (Zhang & Zhu, 2020).

---

[1] Clayton Christensen Institute, "What Is Blended Learning?" https://www.blendedlearning.org/basics/ (accessed July 07, 2018).

It can be explained on a continuum between the combination of hundred percent face-to-face and hundred percent online / computer-mediated delivery of instructions, student engagement, and access to course materials that deliver authentic experiences and storing all the materials for an extended period for the sustenance of adaptive teaching practices, design and development of the future-oriented teaching strategies (Nortvig, Petersen, & Balle, 2018). Blended instructions offer student-centered, teachers facilitated strategy with more adaptability, learning opportunities through collaboration and communication in diverse ways and give chances to solve problems in creative manners (LaBanca, Worwood, Schauss, LaSala, & Donn, 2013).

Days are coming students will continue to demand increased access to technology and flexible nonparallel learning experiences, the driving impetus of which continues to be the pervasive availability and functionality of technology, which would be mobile, ubiquitous, and interactive(Johnson et al., 2016). These emerging developments are conceptualized as a "new culture of learning" where engagement with information happens everywhere, not just in the classroom, and thereby milieu of higher education moves from a stable infrastructure (i.e., learning as the acquisition of a defined collection of knowledge) to a fluid infrastructure. Here teachers' and learners' interaction with knowledge will be continued as a regular phenomenon using technology to create novel applications for existing bodies of knowledge and explore the new streams of knowledge (Smith & Hill, 2019).

Blended learning envisioned a teacher-led instructional method that combines classroom and digital learning, where students have some control over when, where, and at what pace they learn and offers a strategy to help realize this vision. To achieve these outcomes, this method helps tailor instruction to the needs of all students. Here, students engage themselves in learning in a variety of ways: "receiving direct instruction from a teacher in small groups or as a whole class, working in groups collaboratively on a project, rotating between teacher-led instruction and individual work on computers, or working independently on supplemental blended courses. Blended learning can take many forms and is customized to meet student needs.[2]" As a result, it helps to exploit the tremendous potential for enhancing student achievement, preparing graduates for after study engagement, and creating workforce readiness by advancing equity and personalized learning.

## 3. Characterizing the Blended Learning Method

### 3.1. Students Interaction and Engagement

Blended learning has been attributed as a student-centered approach where they can adaptively utilize technology, collaborate, and communicate in multiple ways. They are provided with the environment to solve problems creatively. In the new learning culture, information engagement happens everywhere, not only in the classroom environment. Sometimes information can be shared and exchanged online through classroom discussions, office hours with a professor, lectures, study groups, and papers, by which learning becomes more flexible (Thomas & Brown, 2011).

### 3.2. Stable versus Fluid Infrastructure

In a stable environment, learning systems endeavored to acquire a defined collection of knowledge, whereas, in the dynamics of environmental changes, both teachers and learners interact continually through technological facilities and create novel applications for the existing body of knowledge. Many of the 21st-century faculty members have been attributed as the 'digital immigrants' who stay with modern technology to achieve professional excellence and height of the performance achievement and stay ahead of that task considered an unnerving task (Duderstadt, 1999; Prensky, 2001). Thus, in the fluid infrastructure age, judicious use of technology gives us access to more and better content, communication, and assessment flexibly that turns traditional passive listening teaching experience into an enjoyable transformative learning environment.

---

[2] Colorado Blended Learning Roadmap, July 2017, Produced by Keystone Policy Center.

### 3.3. Continuity of Conversations and Explorations

One of the critical elements of the blended class is the discussion forum that provides online learning opportunities to create and build relationships and facilitate the continuity of conversations and probing (Mozammel et al., 2021). Students can share their thoughts and ideas with every member in the forum and their teacher, prompting through the application of technology and responsive to the review and critique for exploiting the opportunities to learn more by the appealing conversations(Gibbs, Hartviksen, Lehtonen, & Spruce, 2019).

### 3.4. Greater Level of Autonomy and Self-Management

Here faculty provides instantaneous feedback to the students for their submitted works as a means of valuable and personalized learning experience due to their willingness to develop students and engage them in one-to-one meetings directly or virtually. As a result, students in this method develop their self-learning outcomes as standard with greater autonomy and carry bonding for aspiring progress through integrated learning environment and other related modules, co-curricular activities, and other sorts of collaborative and team learning (Bosch & Pool, 2019).

### 3.5. Triggering Sequential Learning

Blended learning provides students unique opportunities to enhance and diversify their knowledge. For example, whenever students post original thoughts and ideas in a discussion forum, that triggers a series of responses, discussions, and conversations in the online learning environment through criticism, affirmation, questioning, and collaboration. Furthermore, limited hours are allocated to the specific module in the institutional schedule in a classroom environment. So, they need to compete in the class with the time constraints, but in blended learning, student interactions and engagement with one another and course content are not restricted to the number of minutes allocated to a class in the schedule. Rather conversations and exploration are continued in the online learning environment (Behzadan & Hsu, 2019).

### 3.6. Multiple Modalities to Address Different Learning Levels of Bloom's Taxonomy

More significant engagements of the students are ensured in the blended classroom, which is more interactive and flexible. Generally, students come in the classroom environment with diverse backgrounds and multiple identities, which again require a multiplicity of learning needs, and blended learning creates the scope of applying bloom's taxonomy conforming to different learning levels (Bloom, 1956). Individualized contents by adaptive and engaging approach as well as embedded assessments and data enhance remembering and understanding levels of the students. However, mini-lessons and teacher instruction increase students' analytical and application-oriented skills (Alrushiedat & Olfman, 2019; Stein & Graham, 2014). However, at the upper levels of the domain, project and group learning helps to improve critical thinking and conceptual evaluations. That is, multimodalities lead towards the development of diversified skills. A simple transmission model of teaching now becomes obsolete to meet the students' needs in the days of significant changes. Thus, contemporarily by blended learning, students build their existing abilities, capacities, and skills in ways relevant to their aspirations, attitude, and aptitude by effectively utilizing the learning development resources and practice-led inquiry (Gosling, 2009).

**3.7. Effects of Benchmarking and Stimulating Collaborative Learning**

Blended learning provides opportunities to learn from the best in the education class from the same levels, contemporary education providers depending on operations and strategic groups. Furthermore, it creates opportunities for collaboration for sharing, disseminating, extending, and innovating knowledge among the various stakeholders of blended learning platforms. Here, wide knowledge sharing removes the limitations in exchanging certain interest-seeking groups (Robinson & Hullinger, 2008; Rübenich, Dorion, & Eberle, 2019).

**3.8. Reducing Mental Stress and Benefits of Permanence**

Blended learning use technology and online learning through which shift contents and instructions to the control of students. It helps to obtain accelerated learning by fundamental redesigning of the instructional model. Thus, through strategic technology integration, learning outcomes are boosted, and talents are leveraged by analyzing real-time student learning and monitoring students' progress. So, methods of instruction are suited to the students evolving needs, and at the same time, it contributes to reducing the students' mental stress and provides benefits of permanence and dynamism in the overall learning process(Philipsen, Tondeur, Roblin, Vanslambrouck, & Zhu, 2019).

**3.9. Creates Opportunities through Systems Thinking and Systems Approach for Accelerating Total Quality Management in the Education**

By focusing on the holistic approach to learning and development, blended learning emphasized creating opportunities through accelerated systems thinking and applying a systems approach to obtain adaptability dynamics. The application of systems thinking helps bring fundamental changes in the academic programs and makes them market-oriented and effective through consistency with micro and macro-environmental changes (Bhuiyan, 2016). Again, education development does not explain and focuses on only the students learning. Instead, it emphasizes total development perspectives, including student learning. Applying the philosophy and practice of blended learning facilitates the application of total quality management with a strong focus on students learning dynamics and innovation. In the days of economic integration, markets focus on continuous opportunities for collaboration and development and create lasting benefits of globalization by integrating changed customers' perceptions into the new strategic directions (Bhuiyan, 2018).

**4. Strategies and Practices of Blended Learning and Teaching at Politeknik Brunei**

Politeknik Brunei, from its inception, has been emphasizing quality education through dynamism and innovation by taking continuous attachment with environmental changes. It is using previous experiences as a learning ground for developing new practices to make them consistent with the market demand and achieving innovation goals of the institution. Towards this direction, this institution is playing very optimistic pro-developmental roles that can be used for further development of the institution in the future and can be a learning ground for other institutions by imitating the strategies for education development through blended learning principles and philosophy. Some of the positive attributes and strategies of Politeknik Brunei are stated below that may deserve the value of learning and imitation in the future:

### 4.1. Pro-developmental Attitude and Extending Support by the Top-Level Management

Optimistic view and insightful look of the top-level management towards the unique environment and relating the analysis for further development of the institution through designing the time consistent strategies are some of the potent factors leading the institution heading towards continuous development. The notable development achieved through blended learning becomes the forefront of further development and milestones for other regional institutions. The institution is continuously learning and getting new experiences by shaping and reshaping its strategies, structuring the functional activities internally, and focusing on the architecture of system thinking. This unleashing tendency of institutional learning becomes the organization's culture, which may be developed as a strategic resource for the institution and accumulated through the institution's intellectual property and once will become a heritage for and hub of the blended learning approach.

### 4.2. Learning Management Systems (LMS) as Platform for Uninterrupted Dialogue

LMS provides multiple ways to tailor instructions according to the needs of students by applying customized strategies. Here students receive instructions directly from the teacher, work on a particular project in a group, learn by rotating between teachers-led instructions to work on computers or independent learning through supplemental blended guidance. By utilizing face-to-face and online learning strategies combined, students can be involved with autonomous learning applying different learning tools simultaneously where interrupted dialogues take place among teachers, students, and systems of learning where students can get the timely need information and learning through the adaptive, contextual environment. So, students in PB get enormous opportunities of learning by using dynamics of opportunities from the relationship and dialogue created by these tripartite parties and systems of the institution. Sometimes, systems and mechanization become alternatives to personal teaching and learning. LMS in PB applies numerous potent learning instruments to ensure effective communication among the participants of learning and e-learning systems of different dimensions, making the learning management system compelling, efficient, dynamic, and catalytic(Bhuiyan, Muliadi, & Hazmi, 2017).

### 4.3. Practices of Team Teaching and Learning

Team teaching and learning is one of the innate characteristics of a PB learning environment. Team teaching provides a parallel learning structure and opportunities for innovation for each member of the team. Members can learn from each other. Principles of collaborative learning can be applied by which the contributions can develop both contents and methods of teaching from the team members. Learning from the team accumulates by increasing the knowledge and upgrade institutional learning as an ingredient of the culture of teaching and learning. From time to time, meetings, intermittent knowledge sharing, mutual discussion, learning from experiences of each other all provide opportunities for continuity of learning and add values to institutional knowledge reservoirs through conceptual and practice development as team culture.

### 4.4. Continuous Professional Development of Teachers and Other Staffs

Continuous professional development is one of the critical determinants of blended teaching and learning practices (Zhang & Zhu, 2020). Therefore, PB provides training opportunities to all faculty members from outside of PB and inside the PB. Thereby all the teachers and staff get the advantage of their career progression. In addition, the culture of at least two presentations by each of the faculty members as one of the critical performance indicators becomes an input for continuous learning and continuous professional development.

### 4.5. Centre for Innovative Teaching and Learning (CiTL) and Innovation Award

The Centre for innovative teaching-learning in the overall organizational structure leads towards the generation of new ideas by the teachers and other members of PB. The open environments where everyone can think and develop new ideas without limiting by any boundaries create the facilitating atmosphere of healthy competition, especially the intergroup competition within the formal structure of PB. The recent introduction of the 'Innovation Award' become one of the motivational factors that may encourage PB family members to think outside of the box individually or with the team for creation and presentation of the new ideas as an immense opportunity for creating new social order in the PB environment and relationship between PB and overall society as well as the global environment.

### 4.6. Strategic Management Unit (SMU) for Providing Policy Supports

The development of SMU is one of the innovative dimensions in PB to provide competitiveness in policy determination and implementation. Strategic management directly contributes to serving the students and society with the excellence of quality and reduction of the cost through efficient operations in every level and cross-functional area of the organizational practices. It creates a causeway between PB and the unique environment by adaptability, change management, surveillance on the overall changes that are coming continuously, and structural setup of the PB itself. The outcomes of this surveillance become the mediating factor in creating competitiveness and adaptability through smooth functioning of the activities in the cross-functional area of operations. SMU provides policy support for adaptability by the market-oriented operational vision and changeability in the all-inclusive structure.

### 4.7. Ensuring Support through Hardware and Software Technology Intervention

It is generally held that any organizations, for its sustainable operations, need to apply updated hardware and software technology to support all operational activities with the least cost, maximum effectiveness, and alignment among the strategy, structure, and people by the mediation of the technological support. PB portal and LMS give 24/7/360-degree operational support to all the stakeholders. All operational phases create opportunities for need-based customized technology-human interface to produce expected operational excellence and services for each stakeholder in the dynamics of structural changes. Furthermore, technology provides connectivity among concerned stakeholders through flexible communication systems, organizing forums, listening sessions, and effective networking systems, sometimes communicating among the virtual team members.

## 5. Factors Limiting the Effective Application of Blended Teaching Philosophy

Despite having multiple benefits of blended learning, it has been found that this approach could not be adequately applied to generate expected benefits due to some limitations. Some constraining factors have been stated below, for which the blended learning approach faces some problems in application.

### 5.1. Unwillingness and Lack of Support from Top Level Management

It is held that any long-term development initiative for an organization should be developed from top-level management of the organization. However, in most cases, developmental initiatives face complexities if higher-level management does not understand it or is unwilling to give wholehearted support for upholding that initiative. Blended learning requires a sagacious plan from the higher-level management along with resource support so that the movement becomes successful without facing any interruptions. However, obtaining success through integrated learning by applying a blended approach requires investment to adapt to the changing needs and make the learning widely accessible to the beneficiaries' group or clients. Top-level management can support the ideas, embrace them, create the proper structure and supportive environment to apply the ideas, make proper plans and apply and execute them, support them through resource provision and finally need to have the courage to accept failure and learn from the experiences.

### 5.2. Lack of Structural Support and Funding

Developmental activities demand a structure having cooperative relationships among the active factors. Structural support may be created by selecting the right people for the jobs to be done, designing proper positions, and involving them in playing the innovative autonomous role to change the existing teaching and learning practices with blended methods. Hybridization of many alternative learning methods consisting of campus and online teaching, resource sharing and dissemination and cooperation, dialogue among the various stakeholder groups, and sustenance of the operations by continuous research and development efforts and renewing the learning from experiences. Thus, sometimes due to lack of structural support and necessary funding, blended learning efforts could not achieve their anticipated and expected benefits.

### 5.3. Dearth of Readiness of the Stakeholders

For being successful through any change, movement organizations should have the readiness to change. That change readiness may require technical aptness from all the participants and willingness to apply the changes throughout system-wide operations by the networking relationships. So, making ready all the participants with blended learning and teaching movement warrants training to the concerned people in the technical and motivational area. Without readiness, the noble job of changing an organization by applying innovative blended learning concepts could not be achieved or interrupted in all its phases of execution.

### 5.4. Gap between Expectation of Teachers and Changed Students Characteristics

A faculty survey conducted in 2009 on student engagement revealed that most instructors do not feel a strong need to use technology although teaching a generation of students who are growing and knowing through the realities of digital culture. In this survey, although 72% of the respondents' report using course management systems (e.g., Blackboard, Learning Studio, Moodle, etc.) but they are reluctant to incorporate other kinds of technology that students commonly use, such as collaborative editing software (such as, Wikis, Google Docs), 16%; blogs, 13%; video conferencing or internet phone chat, 12%; and video games or simulations, 9%. Along with this dismal scenario of faculty's technology use, many of them also do not appreciate the instructional advantage and opportunities that can be created from the effective integration of technology(Werf & Sabatier, 2009).

### 5.5. Blended Teaching Myopia: Focus on Technology, not on Design

Rather than designing the systems with integrating expectations of the stakeholders when more emphasis is given to the technology, it is attributed as technical myopia. Technological myopia creates the obsolescence of products as the market becomes habituated, now expecting better products to serve their needs. For instance, now traditional lecture-based teaching and learning have become outdated. Instead, teaching and learning should be supported with the diversity of information and communication technology support, internet and social media as tools for education, team learning and teaching, virtual team interactions, brainstorming, and other means of innovative learning may complement each other for accelerating the outcome of learning. However, all of these tools should be used objectively; otherwise, technology becomes a fruitless instrument without producing the results or generating the stigma by using technology. Therefore, technology should focus on designing the entire system with result orientation but not as an instrument with any result.

## 6. Framework for Strategy Development and Implementation for Blended Teaching and Learning

A framework may be defined variously. According to businessdictionary.com, a framework may be defined as a "broad overview, outline, or skeleton of interlinked items which supports a particular approach to a specific objective, and serves as a guide that can be modified as required by adding or deleting items" (Businessdictionary.com, 2018). Cambridge English Dictionary defined a framework as a supporting structure around which something can be built or a system of rules, ideas, or beliefs used to plan or decide something (Dictionary.cambridge.org, 2018). In another study, a framework or model has been defined as a systematic representation of some interrelated variables or elements for explaining a phenomenon that serves as a foundation to the cause-effect relationship among the variables relating to that particular phenomenon (Mamun, Bhuiyan, Rab, & Islam, 2000).

By synthesizing the above definitions, a framework may be explained as the abstraction of reality or description of a system and its relevant variables and elements that can be modified based on the environmental changes and considering its contemporariness. In the present study, a framework has been established for strategy development and implementation of blended learning and teaching with particular reference to the academic institutions in the Association for Southeast Asian Nations (ASEAN) region. The initial stage of blended learning strategies requires analyzing both the internal and external environment. Through environmental analysis, the organization could understand its internal strengths and weaknesses and external opportunities and threats. It also helps to understand the needs of the organization and expectations as per the market requirements. After environmental analysis, it is imperatives to develop a wide-angle vision by the participative method.

Accordingly, participation in the development of the vision helps motivate people of the organization in its implementation and creates opportunities for utilization of the employees' potentials. Relevant strategies for blended learning and teaching are readiness to the organization receptive to the changes, clarity of purpose and tasks, technological adaptability, data-driven decision making, feedback and guidance for development, applying flipped classroom, and development of the learning clusters. All the strategies need to be implemented effectively and efficiently. Then, it is relevant to obtain feedback from implementing the strategies and restarting the cycle. Organizations can reenergize the cycle from previous experiences and learning obtained through the implementation of the strategies.

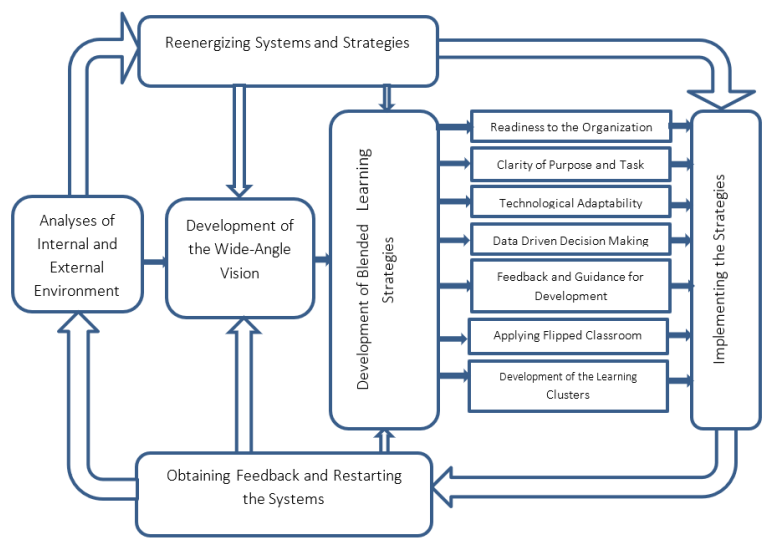

**Figure 1.** Framework for Strategy Development & Implementation for Blended Learning

**6.1. Analyzing Environment and Determining the Needs**

Any strategy before its design is justifiable to analyze the internal and external environmental factors that influence its determination and practical implementation for achieving expected results. The internal analysis will provide knowledge on the strengths, abilities, and weaknesses of the institution. It becomes clear what the institution has and the gaps as per possession, structure, processes, and achievement. In the internal analysis, input factors are teaching, financial, information, and technological resources along with process factors involve method and contents, faculty selection and developing them with appropriate pedagogic skills, providing infrastructural support to meet changing needs of the internal as well as external environment (Bhuiyan, Ahmmed, & Molla, 2009). Moreover, the external analysis will provide organizational adaptability with changing environmental needs emerging from the overall environmental changes. By external analysis, the institution can judge itself in the light of the opportunities and threats. So, by analyzing both factors, the institution can identify the need for designing blended learning strategies in the light of the environmental analysis.

**6.2. Developing Wide Angle Vision and Self-renewal Processes throughout the Organization**

The organization needs to develop a wide-angle vision through the participative method under which blended learning, and teaching philosophy will be guided to achieve the strategic goals. Everyone and group should be adaptive enough for self-renewal of determining and implementing the goals in the organizational settings by considering the impact of surrounding factors on the learning effectiveness and efficiency. Blended Teaching and learning should be designed to achieve diversity of learning and specify the purposes and self-renewal processes that will be used to be adaptive with changes of environment, necessity, the transformation of the characteristics of the learners, and improvement of the technology.

**6.3. Clarity of Purpose and Tasks**

In designing online or offline activities, teachers' approaches and tasks to be achieved by students should be more apparent to them to contribute with their highest potentials. The rationale behind particular approaches, how activities are to be completed, and requirements for solving particular problems should be clarified. Communication regarding teaching and learning methods should be more positive and transparent so that students can understand what to do, when, how, and why within a clear structure. Before assigning any issue or problems, it is better to have a model or example to make it more easily applicable, authentic, and specific to the problems identified and learning objectives. Otherwise, the validity of independent learning will be questioned, and online learning as an instrument will become the source of stress and anxiety.

**6.4. Technological Adaptability with Contemporary Adherence**

Using technology in teaching and learning transforms the role of educators, and the educational process becomes more flexible and student-oriented. For practical and continuing evolutions of strategy development, teachers working with technology involve students to pursue knowledge. Furthermore, adherence to updated technology provides boundless opportunities for teachers to expand their instructions' impact through online sharing of teacher-generated content. Finally, it enables teachers to reach students anywhere flexibly through an online supported remote teaching strategy.

**6.5. Data-Driven Decision Making, Carefully Selected Learning Outcomes and Curricula Development**

To meet the students' needs effectively, Shernon suggested establishing a data-driven culture and using data-driven decision-making in education. To facilitate informed practices, accelerate school performance, and improve students' performance, according to his framework, data-driven decision making should integrate five elements: visionary leadership, collaborative teachers' team for co-learning, set expectations for integrating and modeling through data, developing a data-driven culture and continuous improvement through using technology (Jackson, 2013). In addition, for best teaching practices and learning outcomes, contents and curricula will be developed based on gathering information through continuous assessments and conducting research (Bhuiyan et al., 2009).

**6.6. Prompt Feedback and Instantaneous Guidance for Development**

Concerning the course-related learning outcomes, if faculty provides prompt and detailed feedback to the students regarding their learning, where improvement is needed, it will assist them by providing information for growth and improvement. Blended learning activities, such as discussion, tutorials, online quizzes, assessments, counseling, and other learning activities, should be aligned to the students' contents, learning objectives, and timing. Using lecture, audio-visual instruments, online chat rooms, webcasts, digitized reading documents, social media as information sharing platforms, and other omnipresence instruments create opportunities for on and off-campus studentship through live interaction and prompt feedback to the students and instantaneous guidance for development.

**6.7. Flipped Classroom for Active Learning**

Flip instructions and lectures allow the students to listen to the teacher's lectures at home and apply for tackling complex problems. They can learn through researching, collaborating, crafting, and creating. By providing dynamic and varied learning experiences, students are encouraged to do and debate, make decisions, make questions and be responsive in the flexible learning environment. Here classroom becomes a laboratory for active learning and practices for innovations. Although flipping requires considerable time, commitment, and experimentation from the teachers and skills for planning, filming, and editing quality video presentations that makes the learning enjoyable. In addition to making videos, teachers need to re-conceptualize a new learning and teaching model and rethink utilizing the classroom time to accommodate more active learning with short but attractive video presentations.

**6.8. Developing Learning Clusters**

Learning cluster is the concept of affiliated educational institutions and partnership with the industry players, the result of which is the new type of interaction through social dialogue and collaboration aims to develop regional systems of innovation. Educational cluster contributes to improving conditions for training specialists with different levels of vocational institutions, integration of education with science and industry, and prestige enhancement of highly qualified vocations. An educational cluster is constituting through a group of educational institutions within a specific territory to serve multipurpose roles, such as, to improve educational service, become competitive and interacting providers of necessary factors of industry, equipment, specialized services, utilities, research and development centers to reinforce each other's advantages (Aitbayeva, Zhubanova, Kulgildinova, Tusupbekova, & Uaisova, 2016). Thus, developing an educational cluster becomes an effective means that constitute the partnership and collaborative network between different academic institutions, business players, government, and other support service providers to enhance innovations in educational services, uphold professional standards, and other advantages through mutual benefits interactions.

**7. Conclusive Points and Lessons to be applied by Other Institutions**

Blended learning has been considered a student-centric effective learning strategy that prepares learners for success in a rapidly changing competitive environment. In this direction present study identifies essential characteristics of blended learning methods. The notable characteristics are ensuring students' involvement, flexible structure of learning, continuity of communication, autonomous and self-learning, addressing different levels of learning needs and sequential learning, following benchmark and stimulating collaborative learning, benefits of permanence, and accelerated learning by applying the systems thinking and systems approach. However, some limiting factors have been identified on the way of applying the blended learning approach effectively. However, some limiting factors constrain the practical application of the method. For example, the unwillingness of the top-level management, the gap between teachers' expectations and changed students' characteristics, and blended teaching myopia are some restraining factors hindering the application of blended methods of learning and teaching. By reviewing the strategies and practices of Politeknik Brunei, some key features have been learned: pro-developmental attitude and support of top-level management, learning management system, practices of team teaching and learning, continuous professional development practices, the role of the center for innovative teaching and learning and strategic management unit are notable that contribute significantly in innovating teaching and learning through blended methods, philosophy and practices and obtaining growth through adaptability with environmental changes. However, some suggestions are noteworthy for further development: developing wide-angle vision and self-renewal processes, analyzing environment for needs determination, clarity of purpose and tasks, technological adaptability, data-driven decision making, prompt feedback, flipped classroom, and development of learning clusters for long-term development that can be applied by another similar type of institutions also by considering their environmental distinctiveness.

### 7.1. How to apply

Every institution needs to develop its vision before applying blended learning and teaching to develop knowledge management systems. To develop competitive strategies, top-level management's positive orientation and active support in all dimensions are foremost for an institution. Strategies are to be developed based on analyzing the environment for determining and implementing organization-wide operational plans, and allocating funds is important to support the implementation of the plans. Failure should be taken as learning grounds and based on contingency applying alternative solutions for continuous learning and development. Moreover, finally, imitating the benchmark in the multifunctional area of operations for learning from other institutions may be at the local, regional, national and international level. These are crucial for innovating and adapting learning systems and upgrading the institution for competitive advantage in the long-term perspective and the development of the strategies in a befitting manner.

**References**

Aitbayeva, G. D., Zhubanova, M. K., Kulgildinova, T. A., Tusupbekova, G. M., & Uaisova, G. I. (2016). Formation of education clusters as a way to improve education. *International Journal of Environmental and Science Education, 11*(9), 3053-3064.

Alrushiedat, N., & Olfman, L. (2019). Aiding participation and engagement in a blended learning environment. *Journal of Information Systems Education, 24*(2), 5.

Bath, D., & Bourke, J. (2010). *Getting started with blended learning*. Retrieved from: https://www.griffith.edu.au/__data/assets/pdf_file/0004/267178/Getting_started_with_blended_learning_guide.pdf. [Accessed November 16, 2020].

Behzadan, V., & Hsu, W. (2019). Sequential Triggers for Watermarking of Deep Reinforcement *Learning Policies. arXiv preprint* arXiv:*1906.01126*.

Bhuiyan, B. A. (2001). Training Strategies of Bangladesh Insurance Academy: Exploring the Alternatives for Development. *Insurance Journal, 52*, 65-78.

Bhuiyan, B. A. (2018). Dynamics of consumer behavior in the age of economic integration: Exploring benefits by Bangladesh insurance in the SAARC region. *Insurance Journal, 63*, 76-106.

Bhuiyan, B. A., & Ahmed, S. (2011). The strategic priority for the primary education development in bangladesh: From divergence to the convergence of multidimensional institutions as option. *Teaching Journal of the ooi Junior Academy, 11*(1), 33-44.

Bhuiyan, B. A., Ahmmed, K., & Molla, M. (2009). A theoretical framework for quality assurance in higher education in Bangladesh. *Journal of Business, Society and Science, 1*(1), 27-51.

Bhuiyan, B. A., Muliadi, H. M. B. H., & Hazmi, M. B. H. (2017). *Towards the development of an entrepreneurial academic institution: A case study of Politeknik Brunei*. Paper presented at the International Conference on Business, Economics & Finance (IBCEF), Bandar Seri Begawan: Universiti Brunei Darussalam.

Bosch, C., & Pool, J. (2019). Establishing a Learning Presence: Cooperative Learning, Blended Learning, and Self-Directed Learning *Technology-Supported Teaching and Research Methods for Educators* (pp. 51-74): IGI Global.

Dede, C. (2009). Introduction: A sea change in thinking, knowing, learning, and teaching. In G. Salaway, J. B. Caruso & M. Nelson (Eds.), *The ECAR study of undergraduate students and information technology* (pp. 19-26).

Duderstadt, J. J. (1999). Can colleges and universities survive in the information age. In R. N. K. Associates (Ed.), *Dancing with the devil: Information technology and the new competition in higher education* (pp. 1-26). San Francisco, CA: Jossey-Bass.

Garrison, D. R., & Kanuka, H. (2004). Blended learning: Uncovering its transformative potential in higher education. *The Internet and Higher Education, 7*(2), 95-105.

Gibbs, J., Hartviksen, J., Lehtonen, A., & Spruce, E. (2019). Pedagogies of inclusion: a critical exploration of small-group teaching practice in higher education. *Teaching in Higher Education, 26*(5), 1-16.

Gosling, D. (2009). Supporting Students Learning. In H. Fry, S. Ketteridge & S. Marshall (Eds.), *A handbook for teaching and learning in higher education: Enhancing academic practice* (Third ed., pp. 113-131). New York: Routledge: Taylor & Francis.

Han, F., & Ellis, R. A. (2019). Identifying consistent patterns of quality learning discussions in blended learning. *The Internet and Higher Education, 40*, 12-19.

Jackson, S. (2013). A Continuous Improvement framework: Data-driven decision making in mathematics education. Retrieved from www.dreambox.com.

Johnson, L., Becker, S. A., Cummins, M., Estrada, V., Freeman, A., & Hall, C. (2016). *NMC horizon report: 2016 higher education edition*: The New Media Consortium.

LaBanca, F., Worwood, M., Schauss, S., LaSala, J., & Donn, J. (2013). *Blended instruction: Exploring student-centered pedagogical strategies to promote a technology-enhanced learning environment*. Litchfield: CT: Education Connection.

Mamun, M., Bhuiyan, B. A., Rab, A., & Islam, M. (2000). Promoting entrepreneurship through technological development: A proposed model for developing countries. In P. A. Sinha (Ed.), *Advantage South Asia: Opportunities and Challenges for Management Development* (pp. 420-435). Pune, India: Association for Management Development Institutions in South Asia (AMDISA).

Medina, L. C. (2018). Blended learning: Deficits and prospects in higher education. *Australasian Journal of Educational Technology, 34*(1), 42-56.

Molla, M. S. (2021). Effect of Gender Diversity on the Association between Corporate Sustainability Practices and Financial Performance of Firms. *Finance & Economics Review, 3*(2), 15-31.

Molla, M. S., Hasan, M. T., Miraz, M. H., Azim, M. T., & Hossain, M. K. (2021). The influence of directors' diversity and corporate sustainability practices on firm performance: Evidence from Malaysia. *The Journal of Asian Finance, Economics and Business, 8*(6), 201-212.

Molla, M. S., Ibrahim, Y., & Ishak, Z. (2019). Corporate sustainability practices: A review on the measurements, relevant problems and a proposition. *Global Journal of Management and Business Research, 19*(1), 1-8.

Molla, M. S., Ibrahim, Y., & Ishak, Z. (2019). Relationship between Board Diversity, Corporate Sustainability Practices and Financial Performance of Firms. *Journal of Economics and Sustainability, 1*(2), 22-31.

Mozammel, S., Ahmed, U., & Shakar, N. (2021). COVID-19 and online learning: critical insights for academic achievement. *Elementary Education Online, 20*(4), 1452-1457.

Ngigi, S. K. e., & Obura, E. A. (2019). Blended Learning in Higher Education: Challenges and Opportunities *Handbook of Research on Blended Learning Pedagogies and Professional Development in Higher Education* (pp. 290-306): IGI Global.

Nortvig, A.-M., Petersen, A. K., & Balle, S. H. (2018). A Literature Review of the Factors Influencing E-Learning and Blended Learning in Relation to Learning Outcome, Student Satisfaction and Engagement. *Electronic Journal of e-Learning, 16*(1), 46-55.

Philipsen, B., Tondeur, J., Roblin, N. P., Vanslambrouck, S., & Zhu, C. (2019). Improving teacher professional development for online and blended learning: A systematic meta-aggregative review. *Educational Technology Research and Development*, 1-30.

Prensky, M. (2001). Digital natives, digital immigrants part 1. *On the horizon, 9*(5), 1-6.

Robinson, C. C., & Hullinger, H. (2008). New benchmarks in higher education: Student engagement in online learning. *Journal of Education for Business, 84*(2), 101-109.

Rübenich, N. V., Dorion, E. C. H., & Eberle, L. (2019). Organizational learning and benchmarking in university technology courses: A Brazilian experience. *Benchmarking: An International Journal, 26*(2), 530-547.

Rugova, N. (2021). Social networks as an important part of communication in contemporary trends in adolescents, their impact on their personality and psycho-social behavior. *Technium Social Sciences Journal, 17*, 244-258.

Serrano, D. R., Dea-Ayuela, M. A., Gonzalez-Burgos, E., Serrano-Gil, A., & Lalatsa, A. (2019). Technology-enhanced learning in higher education: How to enhance student engagement through blended learning. *European Journal of Education, 54*(2), 273-286.

Smith, K., & Hill, J. (2019). Defining the nature of blended learning through its depiction in current research. *Higher Education Research & Development, 38*(2), 383-397.

Stein, J., & Graham, C. R. (2014). *Essentials for blended learning: A standards-based guide*, London: Routledge.

Thomas, D., & Brown, J. S. (2011). *A new culture of learning: Cultivating the imagination for a world of constant change* (Vol. 219): CreateSpace Lexington, KY.

Vaughan, N. (2019). A Blended Approach to Teacher Education *Pre-Service and In-Service Teacher Education: Concepts, Methodologies, Tools, and Applications* (pp. 1-22): IGI Global.

Werf, M. V. D., & Sabatier, G. (2009). *The college of 2020: Students*. Washington D.C: Chronicle Research Services.

Zhang, W., & Zhu, C. (2020). Blended Learning as a Good Practice in ESL Courses Compared to F2F Learning and Online Learning. *International Journal of Mobile and Blended Learning (IJMBL), 12*(1), 64-81.